\documentclass[twocolumn]{aastex701}

\begin{document}

\title{Assessment of the 5 August 2026 Falcon 9 Upper-Stage Lunar Impact: Energetics and Crater Dimension Estimation}

\author[orcid=0009-0005-2773-4172]{Mavia Anjum}
\affiliation{Department of Physics, University of Idaho, 875 Perimeter Drive MS 0903, Moscow, ID 83844-0903, USA}
\email[show]{mav.mavia14@gmail.com}

\begin{abstract}
The objective of this study was to characterize the human-made object that impacted the Moon on 5 August 2026, to estimate the dimensions of the crater it produced, and to compare the estimates with the crater subsequently measured by the Lunar Reconnaissance Orbiter (LRO). The impactor was the discarded upper stage of a SpaceX Falcon 9 rocket, which struck the surface near Einstein crater at a velocity of $2.43 \text{ km s}^{-1}$. The kinetic energy and linear momentum of the impactor were calculated, and the crater diameter was estimated using $\pi$-group scaling relationships and measured rocket-body impacts as empirical analogs. The kinetic energy of the impactor was found to be $1.18 \times 10^{10} \text{ J}$, equivalent to approximately 2.8~tonnes of TNT. Compact-body $\pi$-group scaling yielded a final rim-to-rim diameter of 46--57~m, and the empirical analogs yielded 26--27~m, leading to a pre-observation preferred estimate of 25~m within a range of 20--30~m. Between 11 and 12 August 2026, LRO imaged a crater approximately 18~m in diameter and less than 3~m deep. The results revealed that the compact-body scaling overpredicted the observed diameter by a factor of 2.6--3.2, whereas the empirical-analog method overpredicted it by 39--50\%. When the nearest-mass analog was applied as a single crater rather than as a double crater, the estimated diameter was approximately 18~m, in close agreement with the observation. It was found that the hollow and elongated geometry of the stage, rather than a single equivalent bulk density, is the principal factor governing the crater size. This study provides a direct comparison of impact-scaling methods against a measured artificial lunar crater.
\end{abstract}


\section{Introduction}
The exploration of the Moon has resulted in a growing number of human-made objects reaching its surface since the 1960s. Spent rocket stages, defunct spacecraft, and mission-related debris have impacted the Moon both deliberately and accidentally \citep{robinson2023nasa}. These impacts are of scientific interest because, in contrast to natural meteoroid impacts, the mass, velocity, and geometry of the impacting object are approximately known \citep{collins2005earth}. Each fresh crater therefore serves as a controlled cratering experiment in a low-velocity regime that is poorly represented in the natural impact record \citep{holsapple1993scaling, housen2011ejecta}.

The four Apollo S-IVB stages were intentionally directed into the Moon to generate seismic signals for the emplaced lunar seismometers, and produced craters 35--40~m in diameter \citep{robinson2023nasa}. In March 2022, a spent rocket body of uncertain origin impacted the far side near Hertzsprung crater and produced a double crater in which the two lobes measured approximately 18~m and 16~m, with a combined longest dimension of about 28~m \citep{robinson2023nasa}. The double form was attributed to a concentration of mass at each end of the stage, indicating that the crater morphology of a hollow rocket body can retain information about its internal mass distribution.

On 5 August 2026, the discarded upper stage of a SpaceX Falcon 9 rocket impacted the Moon near Einstein crater on the western limb of the near side \citep{kasa2026danuri, yoon2026danuri}. The stage had been left in a chaotic Earth--Moon orbit following the launch of two commercial lunar landers in January 2025 \citep{yoon2026danuri}. Initial observations by South Korea's Danuri (Korea Pathfinder Lunar Orbiter, KPLO) identified the disturbed surface before and immediately after the event \citep{kasa2026danuri}, and between 11 and 12 August 2026 the Lunar Reconnaissance Orbiter (LRO) provided the first high-resolution measurement of the newly formed crater \citep{lroc2026, nasa2026lro, nasa2015lroc}.

The physical and dynamical properties of the impactor were characterized before the event from ground-based observations, which identified an elongated, rotating body and predicted a crater approximately 40~m in diameter \citep{campbell2026physical}. An observational-planning study predicted an impact velocity of $2.43 \text{ km s}^{-1}$ at an angle of approximately $34^\circ$ from the vertical \citep{fernando2026observational}, and the ejecta dynamics were separately modelled \citep{jo2026predicted}. The present study is distinguished from that previous work in that it evaluates crater formation using post-impact energetics and crater-scaling relationships, and it validates the estimates against the crater subsequently measured by LRO. The objective of this study is to determine the kinetic energy of the impactor, to estimate the crater diameter and depth by $\pi$-group scaling and empirical rocket-body analogs, and to assess which method most closely reproduced the observed crater.
\section{The Impactor and Impact Event}
The impactor was the second stage of a Falcon 9 (v1.2 Block 5) launch vehicle. The stage is a thin-walled aluminium-lithium propellant tank, approximately 12.6~m in length and 3.66~m in diameter, with a single engine mounted at one end. Because the propellant tanks are empty at the time of impact, the stage is effectively a hollow cylinder in which most of the mass is concentrated at the engine end. Ground-based characterization further described the object as elongated and rotating \citep{campbell2026physical}, and therefore the mass distribution and impact orientation are expected to be more important than a single equivalent bulk density. A nominal dry mass of 4,000~kg was adopted in this study, with the calculations repeated for the reported values of 3,900~kg and 4,900~kg. The impact occurred on 5 August 2026 at 06:34 UTC at a velocity of $2.43 \text{ km s}^{-1}$. The adopted and observed parameters of the impact are given in Table~\ref{tab:params}.

\begin{deluxetable*}{lll}
\tablecaption{Adopted, estimated, and observed parameters of the 5 August 2026 Falcon 9 lunar impact. \label{tab:params}}
\tablehead{
\colhead{Parameter} & \colhead{Value} & \colhead{Status}
}
\startdata
Object & Falcon 9 upper stage; 2025-010D & identified \\
Mass, $m$ & $\approx 3,900$--4,000~kg (4,000~kg adopted) & adopted \\
Dimensions & $\approx 12.6\text{ m} \times 3.66\text{ m}$ (elongated, rotating) & reported \\
Impact date/time & 5 August 2026, 06:34 UTC & observed \\
Impact velocity, $v$ & $2.43 \text{ km s}^{-1}$ & adopted \\
Impact angle, $\theta$ & $\approx 56^\circ$ from horizontal ($34^\circ$ from vertical) & estimated \\
Target density, $\rho_t$ & $1,500 \text{ kg m}^{-3}$ & assumed \\
Impact site & $19.48^\circ\text{N}$, $266.71^\circ\text{E}$ (near Einstein crater) & observed (LRO) \\
Crater diameter & $\approx 18\text{ m}$ & observed (LRO) \\
Crater depth & $< 3\text{ m}$ & observed (LRO) \\
LRO acquisition & 11--12 August 2026 (NAC) & observed \\
\enddata
\tablecomments{Adopted and assumed parameters were used as model inputs; observed parameters were measured by LRO \citep{fernando2026observational, nasa2026lro, lroc2026}.}
\end{deluxetable*}
\section{Observational Data}
The impact site was imaged by the Danuri orbiter during eight passes at a slant range of $340\text{--}350\text{ km}$, the first acquired approximately 33 minutes before the impact and the remainder over the following hours \citep{yoon2026danuri}. A dark patch with radial streaks, corresponding to disturbed regolith and ejecta, was recorded at the impact point \citep{kasa2026danuri, yoon2026danuri}. Figure 1 presents the post-impact Danuri image alongside an image of the same terrain acquired in 2015 by the Narrow Angle Camera (NAC) of the LRO \citep{nasa2015lroc}, confirming that the feature is newly formed. A short-lived plume containing sodium and lithium was additionally detected through ground-based telescopic observation \citep{vardhan2026spacex}.

\begin{figure*}[htbp]
\centering
\includegraphics[width=\linewidth]{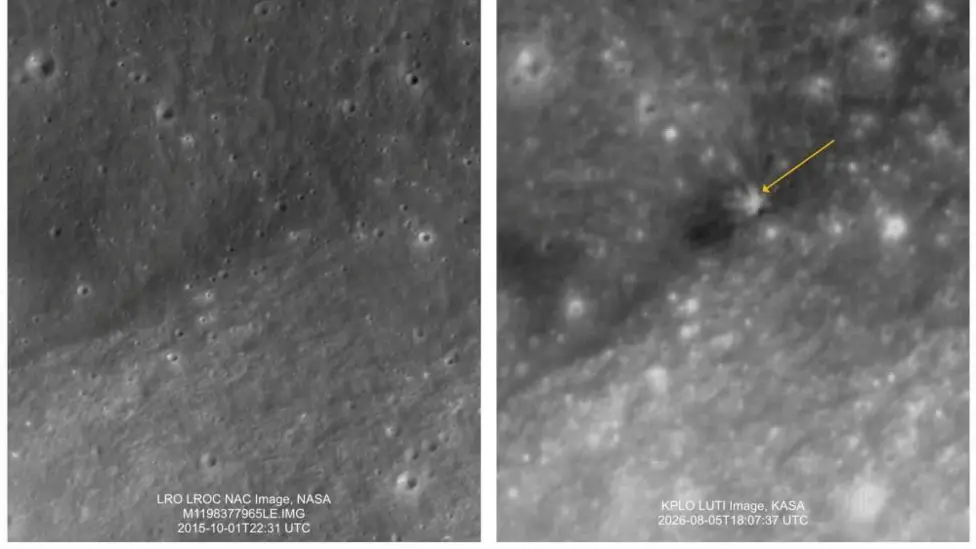}
\caption{Before (left) and after (right) images of the impact site. Left: LRO/LROC NAC frame M1198377965LE, acquired 2015-10-01 \citep{nasa2015lroc}. Right: Danuri/KPLO LUTI image, acquired 2026-08-05 \citep{kasa2026danuri}.}
\label{fig:impact_site}
\end{figure*}
\section{LRO Post-impact Observations}

Between 11 and 12 August 2026, the LRO acquired a series of NAC images of the new crater \citep{nasa2026lro, lroc2026}. The crater was measured to be approximately 18~m in diameter, and its depth was constrained to less than 3~m from the length of the shadow cast on the crater floor \citep{nasa2026lro}. The crater centre was located at $19.48^\circ\text{N}$, $266.71^\circ\text{E}$, at an elevation of 511~m \citep{lroc2026}. Danuri had imaged the site a few hours after the impact and provided refined coordinates to the LRO team, and the pre-impact trajectory prediction was found to be accurate to about one kilometre \citep{nasa2026lro}. Bright and dark rays were observed extending outward from the crater; the darker streaks consist of space-weathered material excavated from the upper 0.46~m of the surface, and the brighter streaks near the rim consist of fresh material from greater depth \citep{nasa2026lro}. A distinctive V-shaped ejecta pattern on the southern side of the crater is consistent with an oblique impact \citep{nasa2026lro}.

\section{Impact Energetics}

The kinetic energy of the impactor was calculated using Equation~(\ref{eq:kinetic_energy}).

\begin{equation}
E = \frac{1}{2} m v^2 \label{eq:kinetic_energy}
\end{equation}

\noindent Here, $E$ is the kinetic energy of the impactor (J), $m$ is the mass of the impactor (kg), and $v$ is the impact velocity ($\text{m s}^{-1}$). 

The TNT-equivalent energy was obtained from Equation~(\ref{eq:tnt_energy}), in which one tonne of TNT corresponds to $4.184 \times 10^9\text{ J}$.

\begin{equation}
E_{\text{TNT}} = \frac{E}{4.184 \times 10^9} \label{eq:tnt_energy}
\end{equation}

\noindent Here, $E_{\text{TNT}}$ is the TNT-equivalent energy (tonnes). The linear momentum of the impactor was calculated as the product of its mass and impact velocity. The calculation was repeated for the reported masses of 3,900, 4,000, and 4,900~kg as a sensitivity analysis.
\section{Crater Scaling}

\subsection{$\pi$-Group Scaling}

An upper-bound transient crater diameter was estimated using the $\pi$-group scaling relationship of \citet{collins2005earth}, given in Equation~(\ref{eq:transient_diameter}), which is based on the framework of \citet{holsapple1993scaling, housen2011ejecta, schmidt1987some}. The compact-body assumption underlying this relationship is not physically representative of a hollow rocket stage, and the result is therefore treated as an upper bound.

\begin{equation}
D_{tc} = 1.161 \left(\frac{\rho_i}{\rho_t}\right)^{1/3} L^{0.78} v^{0.44} g^{-0.22} (\sin \theta)^{1/3} \label{eq:transient_diameter}
\end{equation}

\noindent Here, $D_{tc}$ is the transient crater diameter (m), $\rho_i$ is the impactor density ($\text{kg m}^{-3}$), $\rho_t$ is the target (lunar regolith) density, taken as $1,500 \text{ kg m}^{-3}$, $L$ is the impactor diameter (m), $v$ is the impact velocity ($\text{m s}^{-1}$), $g$ is the lunar surface gravity, taken as $1.62 \text{ m s}^{-2}$, and $\theta$ is the impact angle measured from the horizontal, taken as $45^\circ$ (nominal). 

The final rim-to-rim diameter of a simple crater was obtained from Equation~(\ref{eq:final_diameter}) \citep{collins2005earth}.

\begin{equation}
D_{fr} = 1.25 D_{tc} \label{eq:final_diameter}
\end{equation}

\noindent Here, $D_{fr}$ is the final rim-to-rim crater diameter (m). Because the stage is a hollow body, the equivalent impactor diameter and density were treated as two bounding cases: a solid aluminium sphere of equal mass ($\rho_i = 2,700 \text{ kg m}^{-3}$), and a low-density body in which the mass is distributed over the full stage diameter.

\subsection{Empirical Analog Scaling}

The crater diameter was additionally estimated using measured rocket-body impacts as analogs. In the gravity regime, the crater diameter scales with the impact energy according to Equation~(\ref{eq:empirical_scaling}) \citep{collins2005earth, holsapple1993scaling}.

\begin{equation}
D_{\text{F9}} = D_{an} \left(\frac{E_{\text{F9}}}{E_{an}}\right)^{0.26} \label{eq:empirical_scaling}
\end{equation}

\noindent Here, $D_{\text{F9}}$ is the estimated Falcon 9 crater diameter (m), $D_{an}$ is the measured crater diameter of the analog impact (m), and $E_{\text{F9}}$ and $E_{an}$ are the kinetic energies of the Falcon 9 and analog impactors, respectively (J).

The Apollo S-IVB impacts and the 2022 rocket-body impact were used as analogs \citep{robinson2023nasa}. The 2022 impactor produced a double crater because it carried a secondary payload, whereas the 2025-010D stage carried no secondary payload and a single crater was expected \citep{fernando2026observational}. The 2022 event was therefore applied both as a combined double-crater dimension (28~m) and as a single lobe (18~m).

\section{Results and Discussion}

\subsection{Impact Energetics}

The kinetic energy of the impactor was found to be $1.18 \times 10^{10}\text{ J}$ for the adopted mass of $4,000\text{ kg}$, which is equivalent to approximately $2.8\text{ tonnes}$ of TNT. For the reported masses of $3,900\text{ kg}$ and $4,900\text{ kg}$, the kinetic energy was $1.15 \times 10^{10}\text{ J}$ ($2.8\text{ tonnes}$ of TNT) and $1.45 \times 10^{10}\text{ J}$ ($3.5\text{ tonnes}$ of TNT), respectively. The linear momentum of the impactor was $9.7 \times 10^6\text{ kg m s}^{-1}$. The impact occurred at an unusually low velocity for a natural meteoroid impact, but remained within the hypervelocity impact-cratering regime. At this projectile scale, crater formation is expected to be dominated by gravity-regime scaling.

\subsection{Compact-Body $\pi$-Group Prediction}

\begin{figure}[htbp]
\centering
\includegraphics[width=\linewidth]{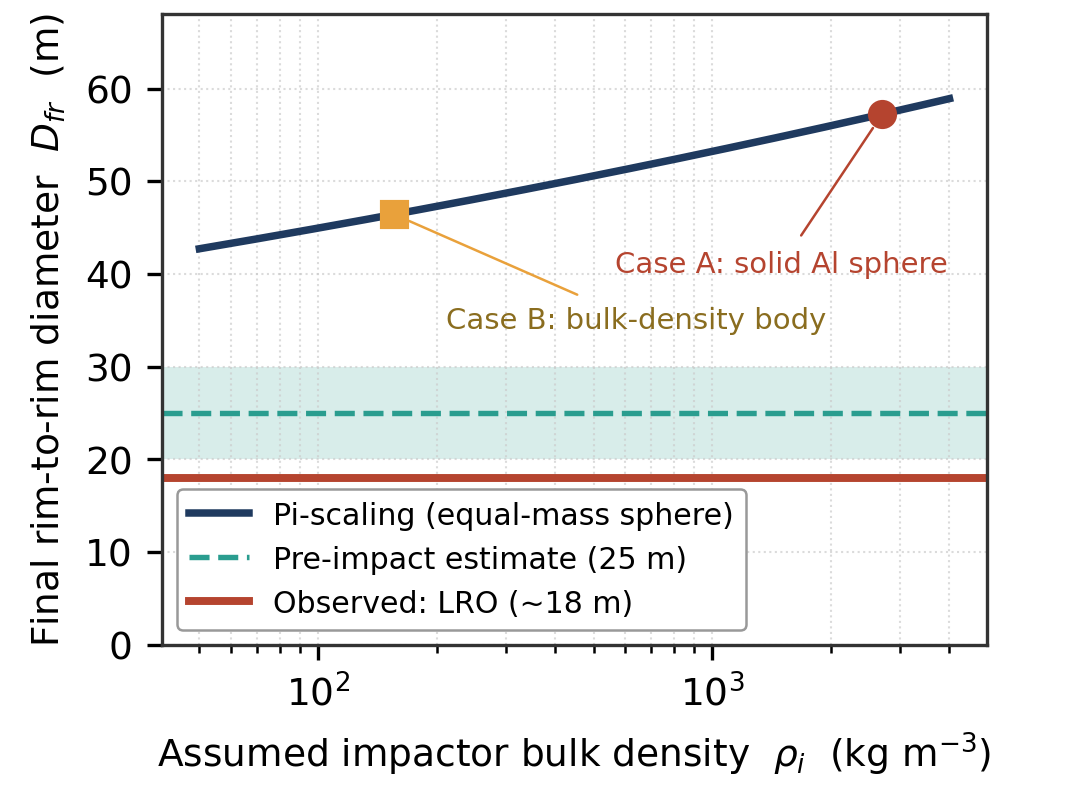}
\caption{Final rim-to-rim crater diameter obtained from Equations~(\ref{eq:transient_diameter}) and (\ref{eq:final_diameter}) as a function of the assumed impactor bulk density (equal-mass sphere, $m = 4,000\text{ kg}$). The two bounding cases lie above the empirically anchored $20\text{--}30\text{ m}$ band (green).}
\label{fig:diameter_density}
\end{figure}

The final rim-to-rim crater diameter was computed as a function of the assumed impactor bulk density using Equations~(\ref{eq:transient_diameter}) and (\ref{eq:final_diameter}), and the results are presented in Figure~\ref{fig:diameter_density}. For the solid aluminium sphere, the final diameter was found to be 57~m, and for the low-density body distributed over the full stage diameter, it was 46~m. Relative to the crater subsequently measured by LRO ($\approx 18\text{ m}$), these values overpredict the observed diameter by 217\% and 156\%, or by factors of 3.2 and 2.6, respectively. This overprediction is attributed to the hollow and elongated structure of the stage, which couples its kinetic energy to the target less efficiently than a compact body of equal mass. For the present event, the compact-body $\pi$-group calculation provides an apparent upper-envelope estimate and substantially overpredicts the observed crater.

\subsection{Empirical Analog Prediction}

A more reliable estimate was obtained from measured rocket-body impacts \citep{robinson2023nasa}. Using the combined $28\text{ m}$ dimension of the 2022 double crater, the estimated diameter was $27\text{ m}$, and using the Apollo S-IVB craters it was $26\text{ m}$. Both estimates overpredict the observed $18\text{ m}$ crater, by $50\%$ and $44\%$, respectively, but far less severely than the compact-body scaling. Because the 2025-010D stage carried no secondary payload, the appropriate analog is a single lobe of the 2022 double crater ($18\text{ m}$) rather than the combined dimension. When the single-lobe dimension was used, the estimated diameter was found to be approximately $18\text{ m}$, in close agreement with the observed crater. The estimate was most accurate when the nearest-mass analog was applied as a single crater.

\begin{figure}[htbp]
\centering
\includegraphics[width=\linewidth]{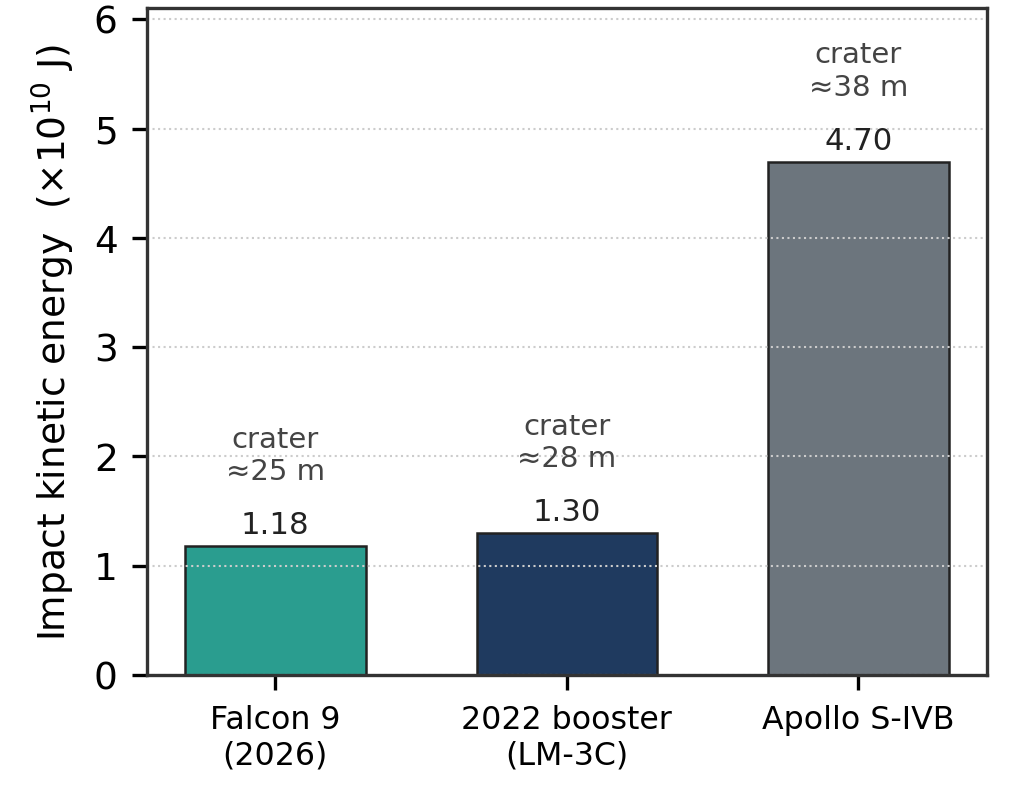}
\caption{Impact kinetic energy of the Falcon 9 stage compared with the 2022 booster and the Apollo S-IVB analogs, with the observed (analogs) and estimated (Falcon 9) crater diameters indicated.}
\label{fig:kinetic_energy}
\end{figure}

\subsection{Effect of Impact Angle and Crater Depth}

The dependence of the crater diameter on the impact angle is weak, because the diameter scales as $(\sin \theta)^{1/3}$. As shown in Figure~\ref{fig:impact_angle}, varying the impact angle from $30^\circ$ to $90^\circ$ changes the estimated diameter by only about $15\%$. The impact angle estimated before the event was approximately $56^\circ$ from the horizontal \citep{fernando2026observational}, and the V-shaped ejecta pattern observed by LRO indicates an oblique impact \citep{nasa2026lro}. The impact angle was therefore not a significant source of uncertainty in the diameter estimate. The crater depth was estimated using a depth-to-diameter ratio of $0.2$, representative of fresh simple lunar craters \citep{melosh1989impact}, and was found to be approximately $5\text{ m}$ for a $25\text{ m}$ crater.

\begin{figure}[htbp]
\centering
\includegraphics[width=\linewidth]{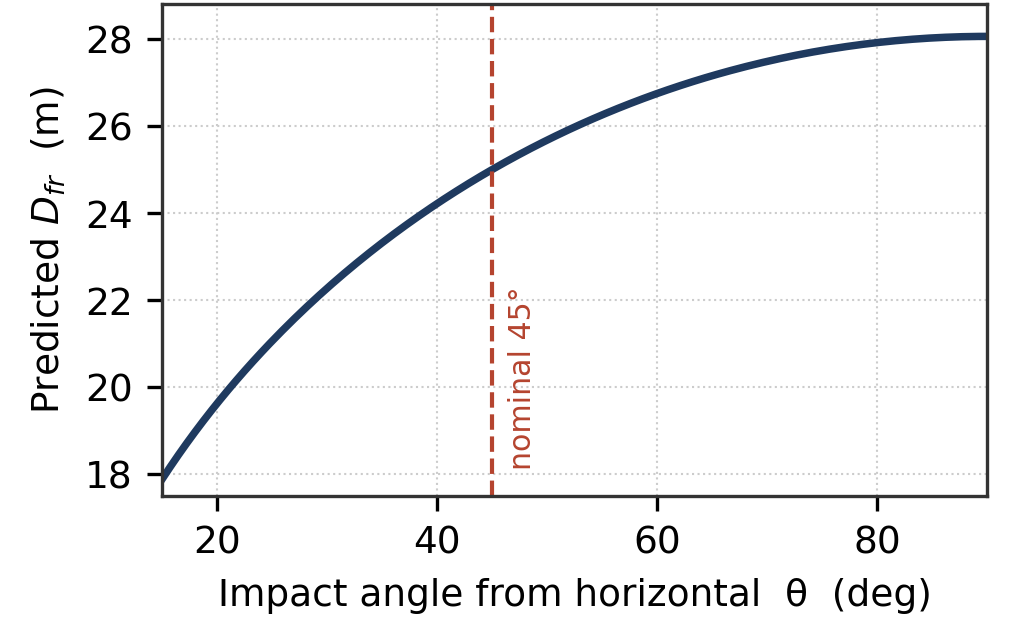}
\caption{Dependence of the estimated crater diameter on the impact angle, normalized to $25\text{ m}$ at the angle of $45^\circ$. The diameter varies by only a few metres over the range of angles.}
\label{fig:impact_angle}
\end{figure}

\begin{figure*}[htbp]
\centering
\includegraphics[width=\linewidth]{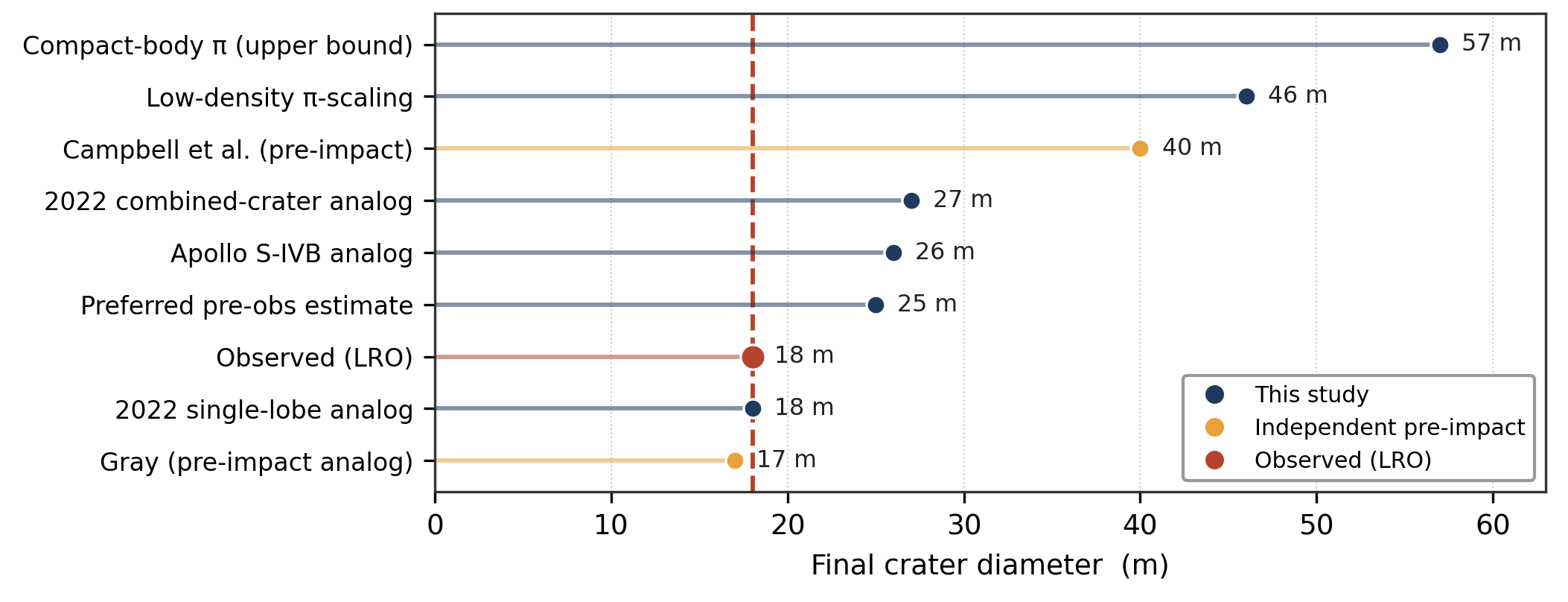}
\caption{Predicted final crater diameters from this study and from independent pre-impact studies, compared with the crater observed by LRO. }
\label{fig:impact_angle}
\end{figure*}

\begin{deluxetable*}{lcccccl}
\tablecaption{Comparison of the Falcon 9 impact with previous artificial lunar impacts. \label{tab:comparison_previous}}
\tablehead{
\colhead{Impact event} & \colhead{Mass ($\text{kg}$)} & \colhead{Velocity ($\text{km s}^{-1}$)} & \colhead{Predicted crater ($\text{m}$)} & \colhead{Observed crater ($\text{m}$)} & \colhead{Ref.}
}
\startdata
Apollo S-IVB stages & $\approx 13,900$ & $\approx 2.6$ & --- & $35\text{--}40$ & \citep{robinson2023nasa} \\
2022 booster (LM-3C) & $\approx 4,000$ & $\approx 2.55$ & --- & 28 & \citep{robinson2023nasa} \\
Falcon 9 (this study) & $\approx 4,000$ & 2.43 & $20\text{--}30$ & $\approx 18$ & \citep{nasa2026lro} \\
\enddata
\tablecomments{Predicted values are from this study; observed values are measured \citep{robinson2023nasa, nasa2026lro}. The 2022 crater is a double crater with lobes of $18\text{ m}$ and $16\text{ m}$.}
\end{deluxetable*}

\subsection{Predicted versus Observed Crater Dimensions}

The predictions of the several methods are compared with the observed crater in Figure~5. The compact-body $\pi$-group scaling ($46\text{--}57\text{ m}$) overpredicted the observed diameter by factors of $2.6\text{--}3.2$. The empirical analogs based on the combined 2022 dimension and the Apollo craters ($26\text{--}27\text{ m}$) overpredicted it by $44\text{--}50\%$, and the preferred pre-observation estimate of $25\text{ m}$ overpredicted it by $39\%$. The lower bound of the $20\text{--}30\text{ m}$ range ($20\text{ m}$) was within $11\%$ of the observation. The single-lobe analog ($18\text{ m}$) and the independent pre-impact analog estimate of Gray ($17\text{ m}$) were the closest to the measured crater, whereas the independent pre-impact estimate of $40\text{ m}$ \citep{campbell2026physical} overpredicted it by $122\%$. The crater depth was also compared. Prior to the LRO observation, a depth-to-diameter ratio of $0.2$ \citep{melosh1989impact} yielded a predicted depth of approximately $5\text{ m}$, whereas the LRO observation subsequently constrained the depth to less than $3\text{ m}$. The empirical-analog method therefore substantially reduced the overprediction relative to compact-body scaling, but the most accurate estimate was obtained when the nearest-mass analog was applied as a single crater.

\subsection{Ejecta Deposit and Disturbed Regolith}

The dark patch recorded by Danuri is larger than the estimated crater. For a final diameter of $25\text{ m}$, the transient crater diameter is approximately $20\text{ m}$, and the continuous ejecta together with the disturbed regolith is expected to extend to two to three crater radii, with ejecta thickness decreasing as approximately the inverse cube of the distance from the rim \citep{collins2005earth, housen2011ejecta, melosh1989impact}. The freshly exposed material is optically darker than the surrounding mature regolith, so the visible scar exceeds the true rim-to-rim diameter; this distinction must be accounted for when a diameter is measured directly from the imaging data. In terms of momentum and seismic energy, the impact was found to be negligible, as the momentum delivered was several orders of magnitude below the value required to perceptibly perturb the Moon.

\subsection{Ejecta Deposit and Disturbed Regolith}

The dark patch recorded by Danuri is larger than the estimated crater. For a final diameter of $25\text{ m}$, the transient crater diameter is approximately $20\text{ m}$, and the continuous ejecta together with the disturbed regolith is expected to extend to two to three crater radii, with ejecta thickness decreasing as approximately the inverse cube of the distance from the rim \citep{collins2005earth, housen2011ejecta, melosh1989impact}. The freshly observed material is optically darker than the surrounding mature regolith, so the visible scar exceeds the true rim-to-rim diameter; this distinction must be accounted for when a diameter is measured directly from the imaging data. In terms of momentum and seismic energy, the impact was found to be negligible, as the momentum delivered was several orders of magnitude below the value required to perceptibly perturb the Moon.

\subsection{Comparison with Previous Artificial Lunar Impacts}

A comparison of the Falcon 9 impact with previous artificial lunar impacts was carried out, and the results are given in Table~\ref{tab:comparison_previous}. The mass of the Falcon 9 stage is comparable to that of the 2022 booster and considerably lower than that of the Apollo S-IVB stages. The observed crater diameter is consistent with a single lobe of the 2022 double crater, which was produced by an impactor of similar mass and velocity. The Apollo S-IVB stages produced larger craters, in agreement with their higher kinetic energy. This comparison is consistent with the interpretation that the crater dimensions of artificial lunar impactors are governed primarily by the impact energy and the internal mass distribution of the impacting object.\\\\\\\

\subsection{Strengths and Limitations}

The strengths of this study are the known identity of the impactor, the constrained impact velocity, the availability of a directly measured post-impact crater, and the direct comparison of theoretical scaling, empirical analogs, and the observed crater. The limitations of the study are the uncertain impact mass and impact angle, the simplified representation of the impactor geometry, the uncertain coupling efficiency of a hollow body, the simplified lunar regolith density, the lack of a direct measurement of the three-dimensional crater morphology, and the approximate nature of the reported LRO diameter and depth. The elongated and rotating geometry of the stage \citep{campbell2026physical}, which is not captured by a single equivalent bulk density, is the most likely explanation for the large crater predicted by the compact-body scaling. \\\\\\

\section{Conclusions}

The aim of this study was to characterize the human-made object that impacted the Moon on 5 August 2026, to estimate the dimensions of the crater it produced, and to compare the estimates with the crater measured by LRO. The kinetic energy of the impactor was found to be $1.18 \times 10^{10}\text{ J}$ for a mass of $4,000\text{ kg}$, equivalent to approximately $2.8\text{ tonnes}$ of TNT. The compact-body $\pi$-group scaling yielded a final diameter of $46\text{--}57\text{ m}$, and the empirical analogs yielded $20\text{--}30\text{ m}$ with a preferred estimate of $25\text{ m}$. Between 11 and 12 August 2026, LRO measured a crater approximately $18\text{ m}$ in diameter and less than $3\text{ m}$ deep. The empirical-analog method substantially reduced the overprediction relative to the compact-body scaling, but still overestimated the measured crater when the combined 2022 dimension was used; the closest estimate was obtained when the nearest-mass analog was applied as a single crater, which reproduced the observed diameter. It was found that the hollow and elongated geometry of the stage is a likely dominant factor governing the crater size, and that compact-body scaling is not appropriate for a hollow rocket stage. This study provides a direct comparison of impact-scaling methods against a measured artificial lunar crater and demonstrates that mass-matched empirical analogs are the more reliable predictor for such impacts.
\begin{acknowledgments}
The author gratefully acknowledges the support, resources, and academic environment provided by the Department of Physics and the University of Idaho Library.
\end{acknowledgments}
\newpage

\bibliography{sample701}
\bibliographystyle{aasjournalv7}

\end{document}